\documentclass[11pt,twoside]{article}
\usepackage{newpasp}

\def\HI{H\,{\sc i}}

\usepackage{epsf}

\index{Oosterloo, T.A.}
\index{Morganti, R.}
\index{Sadler, E.M.}

\begin{document}

\title{HI in Early-Type Galaxies}
\author{Tom Oosterloo \& Raffaella Morganti}
\affil{NFRA, Postbus 2, 9700 AA Dwingeloo, The Netherlands}
\author{Elaine Sadler}
\affil{School of Physics,  University of Sydney, Australia, NSW 2006}

% A concise abstract is recommended.  Enter the text of the abstract in
% between the \begin{abstract} and \end{abstract} commands.  Do NOT
% include the word ``Abstract'' in your text; it is insterted
% automatically. Do NOT  make a paragraph break between \begin{abstract} 
% and the first line of the text of the abstract!  Abstracts are required 
% for all papers.

\begin{abstract}

We briefly discuss the main differences between  the \HI\ properties of
luminous and low-luminosity early-type galaxies. In luminous early-type
galaxies the \HI\ is often irregularly distributed, but in a few cases regular
\HI\ disks, of low surface density,  are seen. In low-luminosity galaxies, the
\HI\ is more often in a disk with high central surface densities. This
suggests a different evolution of the gas in these two groups of galaxies. 

\end{abstract}

\section{Introduction}

Early-type galaxies are generally considered gas poor, but this does not mean
that gas is not important for the evolution of these galaxies.  Many
early-type galaxies, in particular those in the field, do have a long-lived
cool interstellar medium, one that is often very similar to that in spiral
galaxies. The main difference is that they galaxies have less of it,
typically 1-10\% compared to what is present in spirals (e.g.\ Knapp
1999).

Gas content is one of the key elements for understanding the formation and
evolution of galaxies. Early-type galaxies constitute quite a heterogeneous
group of galaxies. Several of their properties (e.g.\ stellar rotation,
isophotal shape, core properties, ionization characteristics of the ionized
gas, star formation history) vary systematically with luminosity and
environment.  These differences can be partly explained by different amounts
of gas present in the formation/evolution. In models for galaxy formation
the gas supply is a key factor. To help in understanding the mechanisms and
processes behind these systematic differences, it may be worthwhile to study
the properties of the neutral hydrogen in early-type galaxies and in
particular whether there are systematic differences of the \HI\ properties as
function of luminosity and environment.

A few dozen \HI\ data cubes are available now of early-type galaxies, most of
them obtained by Van Gorkom and collaborators with the VLA and by us using the
Australia Telescope Compact Array, and we will briefly discuss some of the
systematics of the \HI\ properties of early-type galaxies.

\section{Origin}

One key question is ``where does the \HI\ come from?''. It is generally
thought that the \HI\ is due to a recent accretion/merger event. But is this
the whole story? If one considers the \HI\ detection rates (e.g.\ Bregman et
al.\ 1992), one does indeed see that early-type galaxies that, based on their
optical morphology, are classified as peculiar, are more often detected than
``normal'' early-type galaxies. This does indicate that accretion/merging is
an important factor.  But the statistics also indicate that those early-type
galaxies that show some indications for the presence of a (usually small) disk
component are more likely to be detected in \HI\ (e.g.\ Hogg et al.\ 1993).
This suggests that a subgroup of early-type galaxies has regular \HI\ 
structures.  A third factor is also the environment, although there are no
good statistics available in order to be able to quantify this.

\section{Luminous early-type galaxies}

The range in \HI\ morphology in the  luminous galaxies ($M_B < -19$) is
much broader than that seen in low-luminosity galaxies. In most luminous
galaxies the \HI\ shows an irregular morphology, consistent with the idea that
the gas is accreting onto the galaxy, or is left over from a recent merger
event. Often, these galaxies are found in  small groups. 

However, as suggested by the detection statistics, there are a few luminous
galaxies known that have very regular \HI. An example is NGC 807. This galaxy
is classified as E4 in the RC3, but is rich in \HI. The \HI\ is very extended
(going out to several tens of kpc), and shows very regular kinematics.  Figure
1a shows the position-velocity plot of the \HI\ along the major axis obtained
by us from archival VLA data (Dressel et al.).  Another
example is NGC 3108 (observed by us with the ATCA and the VLA). Also this
galaxy has a \HI\ disk of several tens of kpc in radius. Figure 1b shows the
position-velocity map of the \HI\ along the major axis. The \HI\ disk in NGC
3108 has a central hole. Interestingly, this hole is filled up with ionized
gas that has the same kinematics as the \HI. These data show that the \HI\ in
these galaxies is in a regularly rotating disk, that is basically the same as
disks in spiral galaxies (including the flat rotation curves). In fact, if one
would not know the optical morphology, it would be very difficult to tell from
the \HI\ data alone that these are early-type galaxies.

\begin{figure}
\plottwo{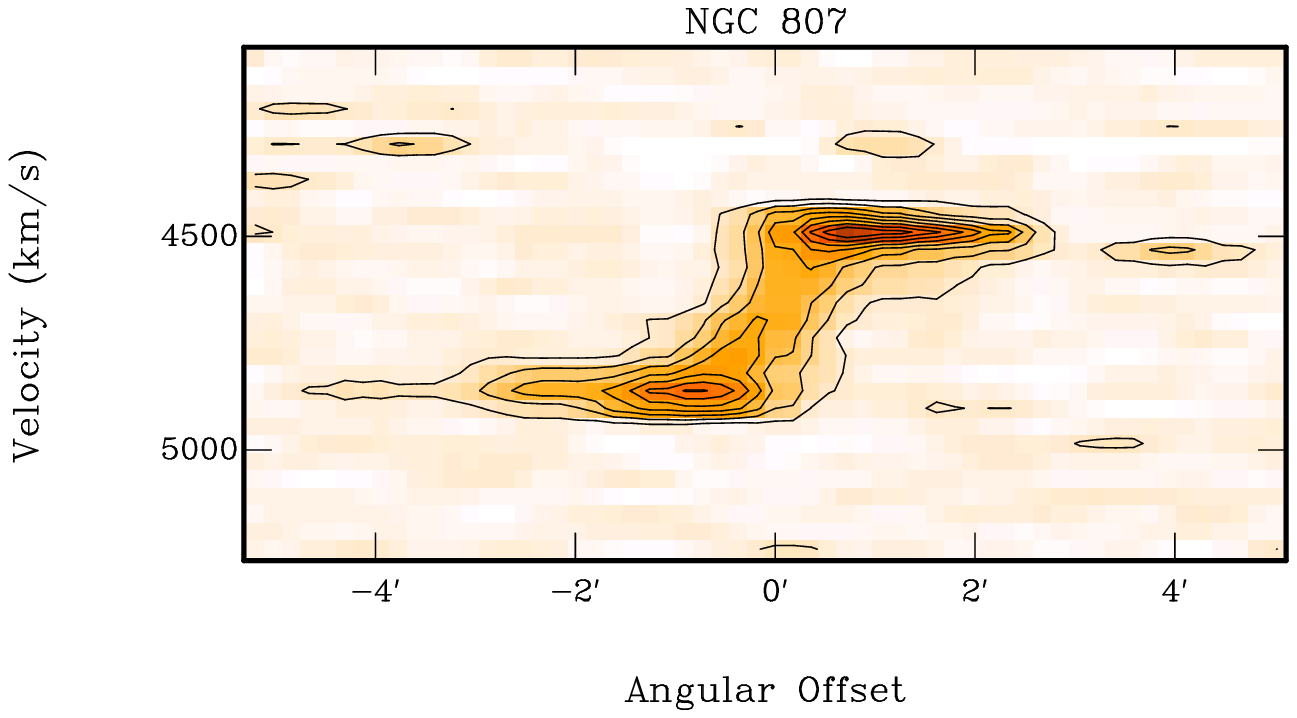}{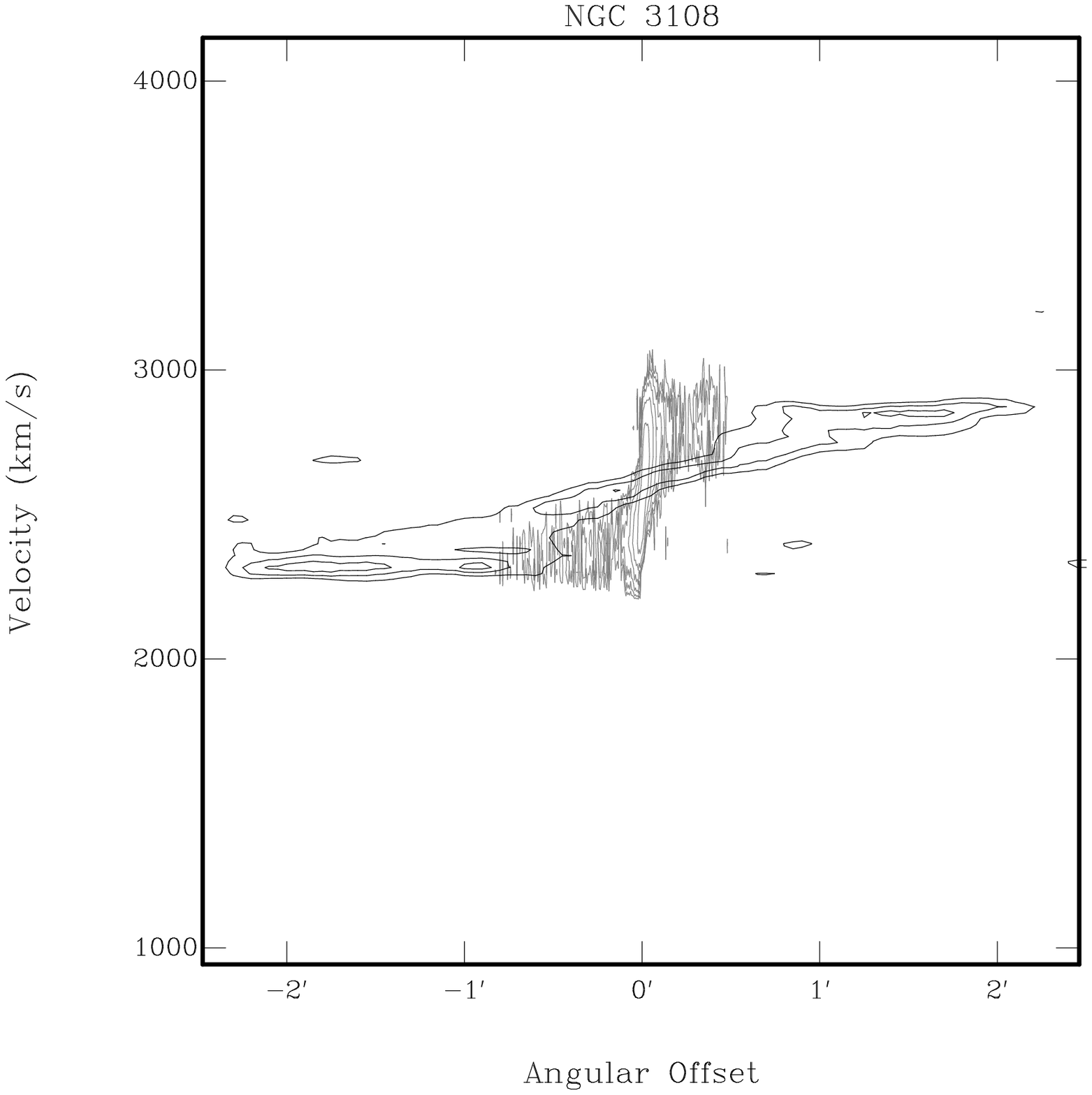}
\caption{{\sl left:} Position-velocity map along the major axis of the E4
  galaxy NGC 807.
{\sl right:} Position-velocity map taken along the major axis of NGC
3108. Black contours show the kinematics of the \HI, the grey contours near
the centre that of H$\alpha$}
\end{figure}

What  distinguishes these \HI\ disks from those seen in spiral
galaxies is that the surface density of the \HI\ is much lower. The peak
surface density is usually below 2 $M_\odot$ pc$^{-2}$ and often the \HI\ 
distribution has a central hole. These low surface densities imply that no
large-scale star formation is occurring and the \HI\ disk will evolve very
slowly. At most a very faint optical counterpart to the \HI\ disk will form.
Some of these galaxies could perhaps be considered as extremely early-type
spirals where the characteristics of the \HI\ are such that an optical disk
failed to form from the \HI\ disk (``disk-less disk galaxies'').

However, the presence of the \HI\ in these galaxies may still be connected to
merging/accretion to some extent. For example, NGC 5266
(Morganti et al.\ 1997) is a minor-axis dust-lane elliptical with a
large amount of \HI\ ($M_{\rm HI}/L_B \sim 0.2$).  Almost all the \HI\ is in a
disk-like structure, with more or less regular kinematics, parallel to the
optical major axis. However, in the centre, a small fraction of
the \HI\ is rotating parallel to the {\sl minor} axis and is coincident with
the minor-axis dust lane. Faint traces of \HI\ connecting the two \HI\ 
structures are also seen.  Most likely, NGC 5266 is a system where a large
amount of \HI\ has been accreted a relatively long time ago and most of the
\HI\ has been able to settle in a disk. Perhaps NGC 5266 is an example of the
scenario suggested by Hibbard and van Gorkom (1996)  that during a
merger of gas-rich galaxies, a large fraction of the \HI\ can be transported
to large radii as tidal tails, while at a later stage this gas
falls back to form a disk.  This would suggest that (some of the) 
early-type galaxies with regular gas disks are very old merger remnants.

\section{Low-luminosity early-type galaxies}

In low-luminosity early-type galaxies the situation is quite opposite to that
in luminous galaxies.  Almost without exception, in low-luminosity early-type
galaxies the \HI\ is in a disk with  regular morphology and kinematics. In a
way, low-luminosity early-type galaxies are also disky in
\HI.   In most low-luminosity galaxies  there is no evidence from the
kinematics that a recent accretion has occurred. This points to a different
accretion history of the \HI. As an example, in figure
3 we give the total \HI\ image and the position-velocity map of the galaxy NGC
2328 ($M_B \simeq -18$). The \HI\ in this galaxy is in a nicely rotating disk,
aligned with the optical body.

Only in some galaxies there is some kinematics evidence that (part of) the
\HI\ may have been accreted recently. For example, in the galaxies NGC 802
(Sadler et al.\ 2000) and NGC 855 (Walsh et al.\ 1990), part of the \HI\ is in
a polar ring-like structure. 

Another characteristic is that in low-luminosity galaxies the \HI\ is very
centrally concentrated and that the central surface densities are much higher
that in luminous early-type galaxies. The central surface densities observed
are typically 6 $M_\odot$ pc$^{-2}$ or higher and significant star formation
is occurring in the central regions. In fact, these galaxies follow the
radio-FIR correlation observed for spiral galaxies, suggesting that the
conditions in the central regions are not too different from those in spiral
galaxies.

\begin{figure}
\plottwo{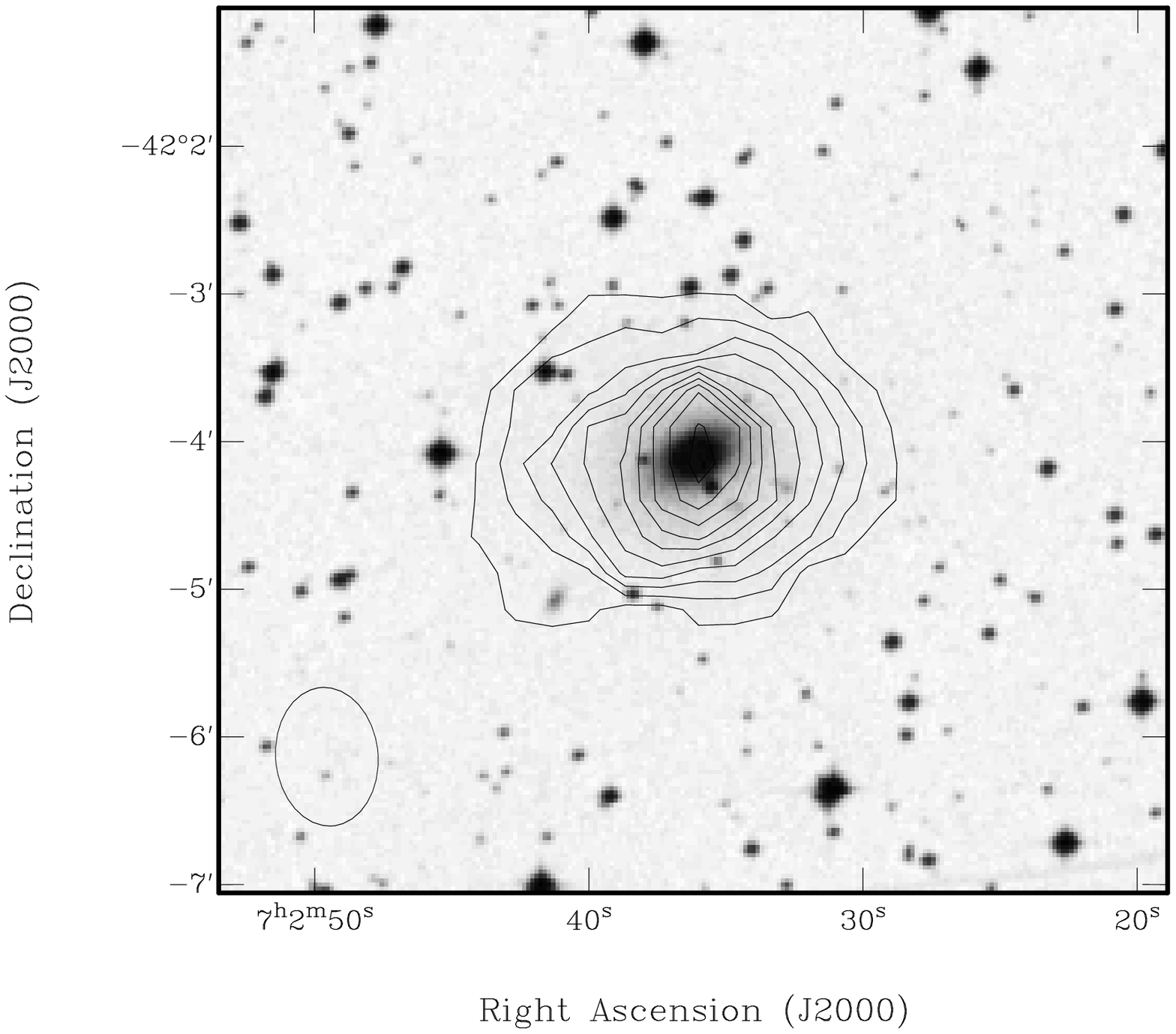}{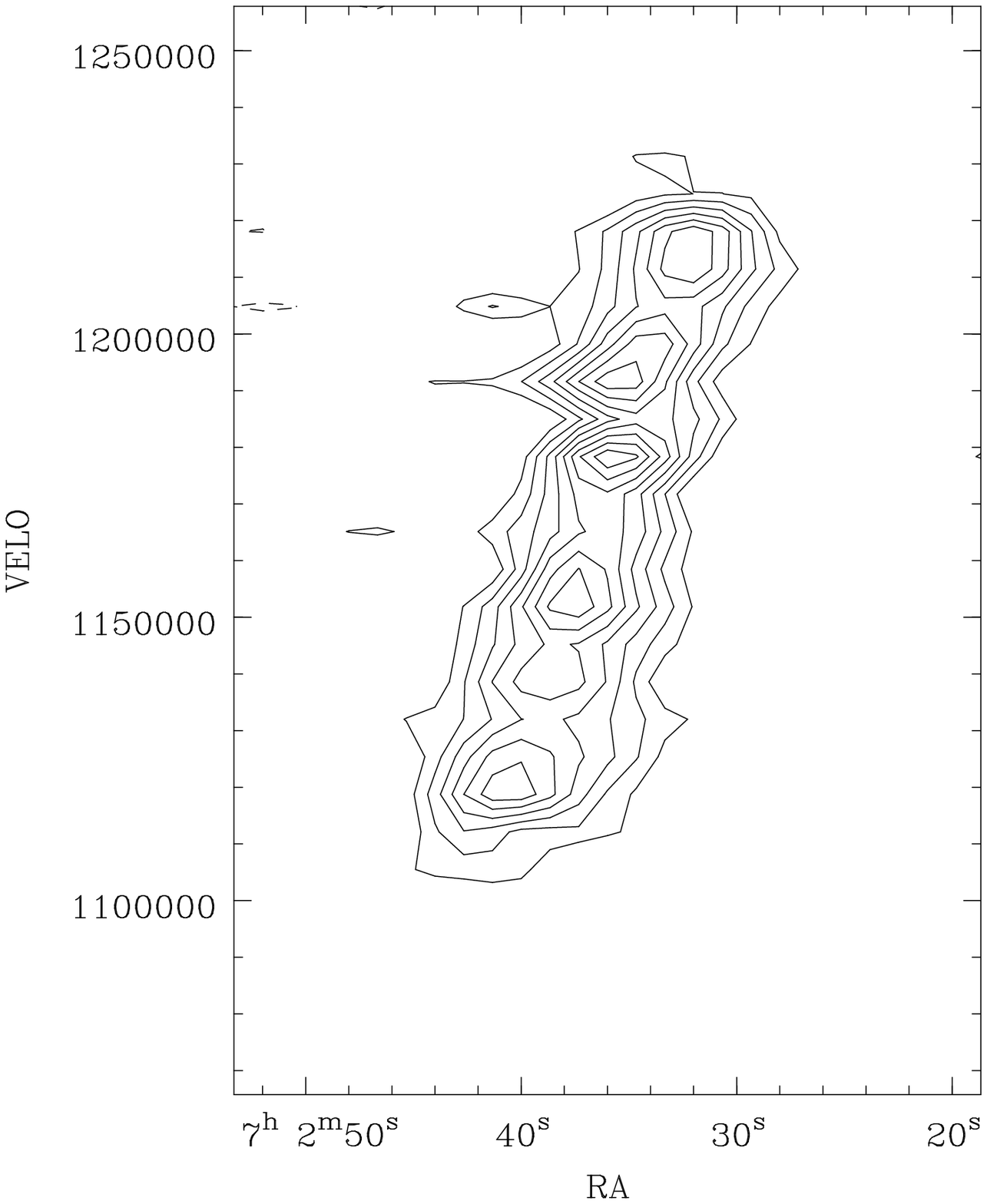}
\caption{{\sl left:} Total \HI\ image of NGC 2328 as obtained with ATCA. The 
optical image  is taken  from the Digital Sky Survey.
{\sl right:} Position-velocity map taken along the major axis of NGC 2328}
\end{figure}

\section{Conclusions}

Systematics differences are observed between the \HI\ properties of  luminous
early-type galaxies and those of low-luminosity early-type galaxies. These
differences are likely to  be connected to  systematics differences between the
two classes of galaxies seen at other wavelengths and could point to 
differences in  accretion history of the gas and  evolution.


\begin{references}
{\small

\reference Bregman, J.N., Hogg, D.E., Roberts, M.S.  1992, ApJ, 387, 484

\reference Hibbard, J. E., van Gorkom, J. H. 1996, AJ, 111, 655

\reference Hogg, D.E., Roberts, M.S., Sandage, A. 1993, AJ, 106, 907


\reference Knapp, G., 1999, in Star Formation in Early-Type Galaxies,
eds. P. Carral and J. Cepa, ASP Conf. Proc. 163, 119

\reference  Morganti, R., Sadler, E., Oosterloo, T., Pizzella, A., 
Bertola, F., 1997, AJ, 113, 937

\reference Sadler, E.M., Oosterloo, T.A., Morganti, R, Karakas, A. 2000, AJ,
119, 1180

\reference  Walsh, D.E.P., Van Gorkom, J.H., Bies, W.E., Katz, N., Knapp, G.R.,
 Wallington, S. 1990, ApJ, 352, 532

}
\end{references}
\end{document}